\documentclass[prb,aps,preprint]{revtex4}
\usepackage{psfig}
\begin{document}
\title{Thermodynamics of $\beta$--amyloid fibril formation}
\author{G. Tiana$^{1,2}$, F. Simona$^1$, R. A. Broglia$^{1,2,3}$ and G. Colombo$^4$}
\address{$^1$Dipartimento di Fisica, Universit\`a di Milano,
Via Celoria 16, I-20133 Milano, Italy.}
\address{$^2$INFN, Sezione di Milano, Via Celoria 16, I-20133 Milano, Italy.}
\address{$^3$The Niels Bohr Institute, University of Copenhagen,
2100 Copenhagen, Denmark.}
\address{$^4$ Istituto di Chimica del Riconoscimento Molecolare, CNR, \\
Via Mario Bianco 9, Milano 20131, Italy.}

\begin{abstract}
Amyloid fibers are aggregates of proteins. They are built out of a peptide called $\beta$--amyloid (A$\beta$) containing between 41 and 43 residues, produced by the action of an enzyme which cleaves a much larger protein known as the Amyloid Precursor Protein (APP). X--ray diffraction experiments have shown that these fibrils are rich in $\beta$--structures, whereas the shape of the peptide displays an $\alpha$--helix structure within the APP in its biologically active conformation. A realistic model of fibril formation is developed based on the seventeen residues A$\beta$12--28 amyloid peptide, which has been shown to form fibrils structurally similar to those of the whole A$\beta$ peptide. With the help of physical arguments and  in keeping with  experimental  findings, the A$\beta$12--28 monomer is assumed to be in four possible states (i.e., native helix conformation, $\beta$--hairpin, globular low--energy state and unfolded state). Making use of these monomeric states, oligomers (dimers, tertramers and octamers) were constructed. With the help of short, detailed Molecular Dynamics (MD) calculations of the three monomers and of a variety of oligomers, energies for these structures were obtained. Making use of these results within the framework of a simple yet realistic model to describe the entropic terms associated with the variety of amyloid conformations, a phase diagram can be calculated of the whole many--body system, leading to a thermodynamical picture in overall agreement with the experimental findings. In particular, the existence of micellar metastable states seem to be a key issue to determine the thermodynamical properties of the system.
\end{abstract}

\maketitle
Amyloid fibers are aggregates of proteins or of fragments of proteins displaying rod--like shape. The mechanism which trigger proteins to leave their biologically active conformation and aggregate is, in general, unknown.

The formation of amyloid fibres \cite{foot1} from monomeric proteins moved to the centre of the scientific stage when it was found to be associated with Alzheimer's disease (AD) and encephalopathies such as bovine spongiform encephalopathy (BSE, "mad cow" disease) in animals and Creutzfeld--Jakob disease (CJD) in humans. The brain of people with the memory disorder  are studded with abnormal structures called plaques, which are made out of amyloid fibers. In this case, the aggregate is built out of a peptide containing from 41 to 43 residues, called $\beta$--amyloid (A$\beta$). This peptide is produced by the action of an enzyme, the $\beta$--secretase which cleaves a much larger protein known as the amyloid precursor protein (APP). This process takes place in everyone, not just people with AD. But people with AD have an increased A$\beta$ production. It has been suggested that this excess $\beta$--amyloid production may lead to fibril formation because, having higher concentration of protein will increase the likelihood that any partially folded intermediates will be able to attach to each other and aggregate.

Recent studies have suggested that oligomeric fibrillation intermediates (protofibrils) rather than fibrils themselves are pathogenic, but the mechanism by which they cause neuronal death remains a mystery.
The possibility that a molecular species other than the amyloid fibril could be pathogenic is testified by the fact that oligomeric species rich in $\beta$--sheet structure (protofibrils) have been found to be discrete intermediates in the fibrillization of $\beta$--amyloid (A$\beta$) {\it in vitro} \cite{landsbury,goldberg}. Furthermore, it has been shown that \cite{lashuel} mutant amyloid proteins associated with familial Alzheimer's and Parkinson's diseases form morphologically indistinguishable annular protofibrils that resemble a class of pore--forming bacterial toxins, that is, toxins which puncture host cell membranes, suggesting that inappropriate membrane permeabilization might be the cause of cell dysfunction and even cell death in amyloid diseases. Observations which suggest that that an intermediate protofibril might be pathogenic and be "detoxified" by conversion to a fibril, is furthermore suggested by the observed lack of correlation between the quantity of fibrillar deposits at autopsy and the clinical severity of Alzheimer's or Parkinson's diseases; the fact that transgenic mouse models of these conditions have disease--like phenotypes before fibrillar deposits can be detected; the fact that non--fibrillar A$\beta$ oligomers are toxic in cell culture\cite{hartley,lambert} and have activity {\it in vivo}\cite{walsh}.

The analysis of the molecular structure of the fibrils, specially of protofibrils, could be useful to understand the aggregation mechanism. Unfortunately, the insoluble and massive character of these systems rules out the possibility of investigating its molecular structure with conventional experimental techniques. X--ray diffraction experiments have shown that these fibrils are rich in $\beta$--structures, and that these structure are perpendicular to the axis of the fibril \cite{kirschner}, whereas the shape of the same peptide in its biological active conformation is an $\alpha$--helix \cite{talafous}. Through electron scanning and atomic force microscopy it has been possible to measure the diameter of the fibrils. The results range from 25$\AA$ \cite{serpell} to $43\AA$ \cite{harper}. Filaments of $90\AA$ in diameter have also been observed. The are built by the assembly of single fibrils \cite{serpell}.

Given this state of affairs, it could prove useful to study the structure of protofibrils and eventually of fibril filaments and the mechanism of formation by molecular dynamics (MD) simulations. However, classical simulations which describe explicitly all atoms of the protein and of the solvent can be carried out over periods of time of only few hundreds of nanoseconds, while the internal motion of a protein and the displacement of its centre of mass takes from microseconds to milliseconds. On the other hand, simplified models which neglect important details of the amino acids and of the solvent are not helpful to study the problem at hand, which strongly depends on the chemical details of the molecular species involved in the process.

The approach which we have preferred to follow consists in performing short full--atom molecular dynamics simulations of the conformational evolution in solvent of the peptides, either isolated or interacting with few other identical peptides, for afterwards use the numerical data obtained to build a thermodynamic model for the formation of larger aggregates.

\section{The starting point: results from MD simulations}

Due to the large computational cost of simulating the dynamics of the whole A$\beta$1--42 amyloid peptide, use was made of a fragment of 17 residues, the A$\beta$12--28 amyloid peptide, which has been shown to form fibrils structurally similar to those of the whole peptide. The amminoacidic sequence of this fragment is:
\vspace{0.3cm}
\centerline{\bf{\fbox{V12}}HHQK{\fbox{LVFFA}}ED{\fbox{V}}GSNK28}
\vspace{0.3cm}
\noindent
where the most hydrophobic residues have been boxed and the first and last amino acids numbered according to their position in the full A$\beta$1--42 peptide.

In order to find the thermodynamically most stable conformations,  we performed 100 ns simulations at 295K and 320K, starting from different conformations of the peptide. Molecular dynamics was implemented through Gromacs software \cite{gromacs}, with explicit representation of all the atoms of the peptide and of the solvent. The equations of motion are solved making use of a leap--frog algorithm in a periodic cell, long--range interaction are treated in the particle mesh Ewald scheme, while temperature and pressure are kept fixed through a weak coupling with an external bath.

Simulations starting from an $\alpha$--helical conformation (cf. Fig. \ref{fig_dimers}, 1H state), which is the crystallografic native conformation of the peptide when embededded in the whole protein, show that the helical features are lost in the first few nanoseconds at both temperatures analyzed. These and other simulations, carried out starting from elongated conformations, lead to the stabilization first of a transient compact state (Fig. \ref{fig_dimers}, 1L state) and then of a hairpin--like state (Fig. \ref{fig_dimers}, 1B state), both characterized by a loop in the region involving residues A21--E22--D23--V24. 

The hairpin-like state is stabilized by the hydrogen bonds, salt bridges and hydrophobic interaction among amino acids in the central region of the peptide (from K16 to D23). This concentration of strongly interacting amino acids makes one suspect that this region is important for the folding of the peptide to its native (helical) state when embedded in the APP protein, perhaps belonging to some local elementary structure (LES) \cite{jchemphys2,jchemphys1,jchemphys3,aggreg} which, assemblying to the other LES into the folding nucleus, is responsible for the correct and fast folding of the whole protein.  This fact would make the effect of the state 1B of the cleaved A-beta peptides even more dramatic than what expected. In fact,not only these peptides aggregate together into amyloid fibrils (as we shall see in the following), but they are likely to interfere with the folding process of entire APP proteins, according to the mechanism discussed in ref. \cite{aggreg}.

Note that the simulations performed in the present work, although computationally demanding (each of the 100 ns run took 6 weeks on a cluster of pentium nodes), are too short to explore the full conformational space of a peptide made out of 17 amino acids. On the other hand, the simulations gave strong indication about the stability of the 1B state and the unstability of the 1H state, indications which are clearly supported by the physical understanding of the factors which stabilize such states (cf. Section IIa) and by the comparison of the predictions of the present model with the available experiments (cf. Section IV). Of course these simulations cannot rule out the presence of other thermodynamically important states (nor can any MD simulation, nonetheless how long, do). However, these other states would have to be rich in $\beta$--structures, as experimentally known \cite{serpell} (cf. also Sect. IV), and neither physical intuition suggests $\beta$--conformations other than those discussed above to be stable, nor it is likely that such conformation would not have emerged from the simulations (the number of different $\beta$--conformations which could be built with a peptide of 17 amino acids being quite small). 

Taking advantage of the knowledge of thermodynamically relevant monomeric states, we performed 30 ns MD simulations of oligomers (dimers, tetramers and octamers), starting from different guesses of oligomeric structures that the monomeric states discussed above can assume. The most stable conformations found from the analysis of the trajectories are displayed in Fig. \ref{fig_oligomers}. It is worth to emphasise that the MD simulations are not meant to explore the full conformational space of such oligomers, but only to search locally around the neighbourhood of the guessed oligomeric conformation. Consequently, the results are strongly dependent on the initial conformations. Such initial conditions are obtained combining the monomeric states discussed above in oligomeric conformations which are physically meaningful in order to optimize the interaction between monomers (e.g., state nC optimizes hydrogen bonds). The main approximation done is, therefore, to assume that optimal oligomeric conformations are spatial arrangements of optimal monomeric conformations, neglecting the fact that the stabilization of monomers into an oligomeric state could strongly change the features of the building blocks. The comparison of the results of the model with the experimental findings, and especially with the structural information known for fibrils and micelles suggests that the approximation is not unreasonable.

\section{Thermodynamical model: infinite dilution approximation}

We first analyze the thermodynamics of monomeric states, which can be viewed as the states relevant in the approximation in which the system is infinitely diluted, so that the (infinite) translational entropy prevent monomers from binding together into aggregates.

\subsection{Free energy of monomeric states}

The interest in the $\alpha$--helical state (Fig. \ref{fig_dimers}, "1H") relies on the fact that, although this conformation, according to our simulations,  is not stable,  it is the crystallographic structure for this sequence when forming part of the A$\beta$1--42  wild--type protein under biological conditions. 

The internal energy is calculated by averaging the potential energy (i.e., Coulomb and Van der Waals contribution) in a 2 ns time interval at 295K, short enough for the fragment to remain in the elical conformation. We obtain \cite{foot2} $E_{1H}=-887$. 

In the same MD--short trajectory,  the entropy of the sidechain is also calculated. Since the motion of the sidechain takes place on a time scale which is much shorter than that of the simulation, and since the backbone is essentially fixed during this time interval, it is possible to assume that the sidechain is at equilibrium. Thus, one can obtain the free energy from the logarithm of the partition function,
\begin{equation}
F^{sidechain}=-T\log<\exp(\frac{-E}{T})>.
\end{equation}
The brackets $<>$ indicate the thermodynamical average, which in this case gives rise to the partition function, and is set equal to the time average by virtue of the ergodic theorem. Subtracting from the free energy the average interaction energy, one obtains the entropy associated with the motion of the sidechain
\begin{equation}
(T{S^{side}})_{1H}=-88.6.
\end{equation}

Since the state 1H of A$\beta$12--28 corresponds to a unique conformation of the backbone (the calculated RMSD being lower than 2$\AA$), separated from the other conformations by a sharp transitions (for details see ref. \cite{fabio}), the entropy associated with the backbone is set equal to zero. 

The solvation free energy $\Delta G_{solv}$, i.e. the difference in free energy \cite{foot3} associated with the interaction between the solvent and the A$\beta$12--28 peptide in an elongated (random) and in the 1H conformation is the result of two effects: a decrease of the free energy due to the fact that the hydrophobic  side chains become partially hidden to the solvent and an increase of the free energy due to the fact that the new compact 1H state some hydrogen bonds between hydrophobic residues and water present in the unfolded conformation are not allowed in the folded conformation. The quantity $\Delta G_{solv}$ for the 1H state cannot be obtained directly from the simulations because, within the short time over which the MD simulations are carried out, the motion of the solvent molecules cannot be approximated as at equilibrium. Instead, we obtain the numerical value of $\Delta G_{solv}$ from the FOLD-X \cite{foldx} approximated force field, which gives to each residue in contact with the solvent a contribution to the free energy proportional to the exposed surface to the solvent and the degree of hydrophobicity  of the amino acid. One obtains
\begin{equation}
(\Delta G^{solv})_{1H}=62.3
\end{equation}

The total free energy for the state 1H at 295K is obtained from the sum of all the contributions discussed above and gives
\begin{equation}
F_{1H}=-736.1.
\end{equation}
We choose the state 1H as reference for the following calculations, and accordingly we set $\Delta F_{1H}=0$. Note that the choice of the free energy of the reference state, provided that it is the same for all states, is purely conventional, and has no consequence on the following calculation.

The free energy of the hairpin--like state 1B calculated in a similar way, giving $E_{1B}=-881.1$, $TS_{1B}^{side}=-100.0$ and $(\Delta G)_{1B}=45.8$. Unlike the previous case, the state 1B displays a non--zero entropy associated with the backbone. This is because the harpin defines a flexible plane which can bend perpendicularly to the direction of the hydrogen bonds (cf. Fig. \ref{fig3new2}).
Since a direct calculation of this contribution directly from MD simulations is out of reach, it was estimated by means of a simple model. If each pair of residues connected by hydrogen bonds gives rise to $\eta$ conformations, the total number of conformations that an hairpin built out of $N$ residues can assume is $\eta^{(N-1)/2}$. Consequently
\begin{equation}
\label{ent-1b}
S^{chain}_{1B}=(N-1)/2\cdot\ln\eta.
\end{equation}
The value of $\eta$ is not precisely known, but it can be estimated in keeping with the fact that in an unconstrained polypetide each residues can assume $\approx 9$ conformations, and that, since the residues in an hairpin are bound by hydrogen bonds, they will display approximately half of the degrees of freedom of the unconstrained chain, so that $\eta=9^{1/2}=3$. Within this approximation $TS^{chain}_{1B}=21.6$ at 295K, leading to $F_{1B}=-756.9$, and, with respect to the reference state 1H, $\Delta F_{1B}=F_{1B}-F_{1H}=-20.8$.

Another state which show a sizable stability is the state 1L. As in the case of the 1H state, this is associated with a unique backbone conformation, so that $S_{1L}^{chain}=0$. From the simulations one obtains $E_{1L}=-1055.7$, $TS^{side}_{1L}=-237.4$ and $\Delta G^{sol}_{1L}=66.2$, the total free energy summing to $F_{1L}=-752.1$ ($\Delta F_{1L}=-16.0$ with respect to the reference state 1H).

In order to account correctly for the thermodynamics of the system, one has to include in the calculations also the set of denaturated states ("1U"). Although not displaying low interaction energy, they display a large entropic contribution, and consequently become more relevant the higher is the temperature.
Also in this case, it is not possible to obtain the free energy from MD simulations, which are by far too short to provide an exaustive search in conformational space. Instead, we will describe the states of type 1U with the help of the Random Energy Model \cite{derrida}. Assuming that the interaction between residues in denaturated states are random and uncorrelated, and that the chain is long enough that the Central Limit Theorem holds, the energy distribution for states of type 1U (density of states) is
\begin{equation}
\label{poe}
g(E)=\gamma^{N} k^{-1}\exp\left[-\frac{(E-N_c\epsilon_0)^2}{2N_c\sigma^2}\right],
\end{equation}
where  $N$ is the length of the chain, $\gamma$ is the number of conformations per residue, $k=(2\pi N_c\sigma^2)^{1/2}$ is the normalization constant, $N_c$ being the number of two--body interactions in the chain, $\epsilon_0$ and $\sigma$ are the average and the standard deviation of the two--body interactions.

In the present case, $N=17$ and we shall assume that the number of interaction contacts $N_c$ is constant for all conformations in the state 1U and equal to the average value found in the simulations, that is $N_c=16$.

From Eq. (\ref{poe}) one finds that the entropy of the chain, defined as $\ln g(E)$ (aside from an overall constant) is given by
\begin{equation}
\label{entropy_rem}
S^{chain}(E)=N_c\ln\gamma'-\frac{(E-N_c\epsilon_0)^2}{2N_c\sigma^2},
\end{equation}
in the case where $E>E_c\equiv N_c\epsilon_0-N_c\sigma(2\log\gamma')^{1/2}$, with $\gamma'\equiv\gamma^{N/N_c}$. If $E<E_c$ the argument of the exponential in Eq. (\ref{poe}) diverges to $-\infty$ linearly with $N$ and there are essentially no states with energy below $E_c$. Making use of the relation $\partial S/\partial E=1/T$ one also finds
\begin{equation}
<E>=-\frac{N_c\sigma^2}{T}+N_c\epsilon_0.
\end{equation}

In order to calculate the numerical values of $\epsilon_0$ and $\sigma$, we have identified the states 1U as those parts of the trajectories corresponding to conformations structurally different \cite{foot4} from state 1B. From the analysis of the trajectories \cite{fabio} it emerges that these states are separated from state 1U by strong variations in all thermodynamical quantities, and consequently we can interpret 1U as a metastable state, separated from the others by a free--energy barrier. The associated Boltzmann--like energy probability $p(E)\equiv g(E)\exp(-E/T)/Z$,  the entropy being that found in Eq. (\ref{entropy_rem}). This gives
\begin{equation}
\label{concavity}
\frac{\partial^2}{\partial E^2}\ln p(E)=-\frac{1}{N_c\sigma^2}.
\end{equation}
Finding the concavity of $p(E)$ and the value of $E_c$ from the MD simulation, from Eq. (\ref{concavity}) and from the definition of $E_c$ one finds $\sigma=12.2$ and \cite{foot5} $\Delta\epsilon_0\equiv \epsilon_0-F_{1H}/N_c=\epsilon_0+46=14.8$.

From the analysis of the MD trajectories we have also found that $T\Delta S^{side}_{1U}=+1.5$. The solvation free energy has been calculated with FOLD-X as the average of the solvation energies of ten conformations picked at random from the simulations. The total free energy is found as logarithm of the partition function. This is calculated integrating the free energy over all possible energies associated with the state 1U, that is from $E_c$ to $+\infty$. This gives
\begin{eqnarray}
\exp(\Delta F_{1U})&=&\exp\left(N\log\gamma+\Delta S_{1U}^{side}-\frac{\Delta G^{sol}_{1U}}{T}-\frac{N\Delta\epsilon_0^2}{2\sigma^2}+\frac{1}{2}N_c\sigma^2\left(\frac{\Delta\epsilon_0}{\sigma^2}-\frac{1}{T}\right)^2\right)\times\nonumber\\
&\times&\frac{1}{2}\left[1-\mbox{erf}\left( \frac{\Delta E_c+(N\sigma^2(\Delta\epsilon_0/\sigma^2-T^{-1}))}{(2N_c\sigma^2)^{1/2}}\right)  \right].
\label{pippo}
\end{eqnarray}
At a temperature of 295K this relation gives $\Delta F_{1U}=-13.4$. The results discussed in the present Section are summarized in Table \ref{newtab1}.

\subsection{Population of monomeric states}

From the knowledge of the free energy of all the thermodynamicall relevant states, it is possible to calculate their equilibrium population. The partition function reads
\begin{eqnarray}
Z&=& 1+\exp(-\Delta F_{1B}/T)+\exp(-\Delta F_{1L}/T)+\exp(-\Delta F_{1U}/T).
\end{eqnarray}
 
The most populated state at 295K results to be the $\beta$--hairpin (1B), whose relative probability of occupation is
\begin{equation}
p_{1B}=\frac{\exp(-\Delta F_{1B}/T)}{Z}=0.91,
\end{equation}
while the 1H state displays a negligible relative probability.

A systematic analysis of relative state populations as a function of temperature poses some problems. First, an analitic expression of the solvation energy with respect to temperature is not available. Moreover, extrapolating at high temperature thermodynamical quantities found at room temperature is not straightforward. For instance, as the temperature rises, the solvent undergoes a phase transition from liquid to vapour, and the extrapolation becomes meaningless. Moreover, the average interaction energy and the side chain entropy are only calculated at 295K and 320K.

In any case, between 273K and 373K the solvation free energy displays variations below 30\% \cite{creighton} and one can attempt an extrapolation. The free energy associated with the state 1U, calculated from Eq. (\ref{pippo}), is showed in the inset of Fig. \ref{mathe} and is well approximated by $\Delta F_{1U}=-16.5T+29.0$. The relative population of the harpin state 1B is then
\begin{equation}
p_{1B}=\frac{\exp(\frac{10.4}{T}+4.28)}{1+\exp(\frac{164.1}{T}-59.5)+\exp(\frac{10.4}{T}+4.28)+\exp(\frac{-29.0}{T}+16.5)},
\end{equation}
and is displayed in Fig. \ref{mathe}.

The fact that $p(1B)$ increases with temperature is consistent with the experimental data obtained by circular dichroism \cite{serpell} and is not trivial, as, in general, physical systems loose symmetry as the temperature is increased. This particular behaviour is consequence of two facts: 1) the denaturated state 1U has a free energy quite larger than the state 1B, so that it starts competing with the state 1B only at high temperatures, 2) the state 1B displays a conspicuous entropy term associated with thebackbone degrees of freedom on the plane defined by the hairpin (cf. Fig. \ref{fig3new2}), term which is responsible for the decreasing the free energy as temperature is increased.

\section{Thermodynamical model: aggregation}

A solution of $\beta$--amyloid peptides contains a large number of monomers (i.e., peptides) which can interact and give rise to conformations which are more complicated than those discussed in the infinite dilution limit. Typically, the concentration of peptides is of the order of $100\mu$M ($=10^{-4}$ mol/l), the volume of a cuvette is $V=1\; ml$, so the number of monomers is of the order of $10^{16}$.

\subsection{Free energy of the aggregates}

The free energy of the multimeric states is calculated in a way similar to that of monomers (see previous Section), by including the interaction energy between different monomers and the entropy associated with the translational and rotational degrees of freedom of the monomers. The inter--monomer energy is calculated directly from the MD simulations (in the same way as the intra--monomer energy), while the new entropic term is approximated as
\begin{equation}
\label{strasl}
S^{rot-trasl}(n)=\ln\left[\frac{V-(n-1)v}{v}\right]+(n-1)\ln\left[\frac{4\pi^3}{\Delta\omega}\right],
\end{equation}
where $n$ is the number of monomers present in the volume $V$. The first term is associated with the translational motion of the centre of mass of each monomer, whose Van der Waals volume is $v=2$nm$^3$ \cite{creighton} in a volume $V$, while the latter term accounts for the rotational degrees of freedom. The quantity $\Delta\omega$ is the product of the amplitudes of the Euler angles available to each pair of bound monomers and it takes a value of the order of $\Delta\omega\approx 8\cdot 10^{-3}$ (see Appendix). Notice that in the first term of Eq. (\ref{strasl}), the term $(n-1)v$ is not negligible as compared to $V$ only at concentrations larger than $10$M, concentrations which are non--biological (experimental concentrations are of the order of $100\mu$M). Consequently, we can assume $V\gg (n-1)v$, and approximate Eq. (\ref{strasl}) as $S^{rot-trasl}(n)\approx\sigma^{rot-trasl}n+\sigma^{rot-trasl}_0$, where we define $\sigma^{rot-trasl}=9.6$ and $\sigma^{rot-trasl}_0=38.7$.

We define as reference state for an aggregate of $n$ monomers the set "nH" of conformations composed on $n$ non--interacting monomers in helical state (1H), whose free energy is $F_{nH}=nF_{1H}-TS^{rot-trasl}$.

The states relevant for aggregations are chosen on the basis of the data resulting from the MD simulations discussed in Section I. In particular, the free energies associated with some of the states displayed in Fig. \ref{fig_oligomers} are listed in Table \ref{table_2f}. The state displaying two interacting helices (labelled "2K" to distinguish it from the state "2H" displayin two non--interacting helices) has a free energy much higher than the other states. Consequently, we will not take it into account, nor the multimeric states built starting from it. We will indicate each kind of aggregate with a number, which indicate how many monomers which build out that aggregate, and a label, which indicates its shape. For example, 18$P_2$ means an aggregate built out of 18 monomers arranged as indicated in Fig. \ref{fig_oligomers}.

The states nP$_1$, nP$_2$, nP$_4$ and nC (see Fig. \ref{fig_oligomers}) are those which, for large $n$, resemble (proto) fibrils. Under the assumption that each layer of the fibril (each section (layer) composed of 1 monomer in the case of states nP$_1$ and nC, of two monomers in the case of nP$_2$ and of four monomers in the case of nP$_4$) interacts only with its nearest neighbours and, under the approximation that the solvation free energy is additive, the total free energy for a given kind of aggregate is linear functions in $n$. The result for the state of kind $nk$ (where $nk$ can be nP$_1$, nP$_2$, nP$_4$ and nC) is
\begin{equation}
\label{free_fibrils}
\Delta F_{nk} = n\epsilon^{mon}(k)+n\epsilon^{int}(k)+\epsilon^{int}_0(k)-nT\sigma^{side}(k)-nT\sigma^{chain}(k)+n\rho^{solv}(k)+TS^{rot-trasl}.
\end{equation}
The quantities $\epsilon^{mon}(k)$, $\epsilon^{int}(k)$, $\sigma^{side}(k)$, $\sigma^{chain}(k)$ and $\rho^{solv}(k)$ are the energy densities listed in Table \ref{free_dens}, obtained from linear extrapolation of the data listed in Tables \ref{newtab1} and \ref{table_2f}. In particular, $\epsilon^{mon}(k)$ is the density of internal energy and $\epsilon^{int}(k)$ is the density of interaction energy between monomers. The translational entropy is that given in Eq. (\ref{strasl}). The quantity $\epsilon^{int}_0(k)$ takes into account that a single layer does not produce any interaction energy (i.e., the interaction energy of the state nP$_1$ is proportional to $n-1$, not to $n$).

Other thermodynamically important states are those which display the hydrophobic head of the 1B conformation towards the centre of the aggregate, in a way similar to lipidic micelles (see Fig. \ref{fig_oligomers}). They are less ordered than fibrils, but gain free energy by hiding the hydrophobic residues from the solvent.

The density of internal energy per monomer is not constant, but increases when going from dimeric to octameric micelles (being 14 for 2M, 24 for 4M and 49.6 for 8M). This could be due to the increasing difficulty of optimizing the contacts between amino acids in a larger, more frustrated system. Since we are interested in extrapolating the internal energy for large $n$, we will use the largest value, setting $E_{nM}=49.6n$. The inter--monomer interaction energy, the solvation free energy and the sidechain entropy of the dimeric, tetrameric and octameric micelles are displayed in Fig. \ref{micella_ene}. While $E^{int}$ is almost linear with $n$, $\Delta G^{solv}$ and $TS^{side}$ are only approximately linear. This is consistent with the fact that the solvation energy is not purely additive \cite{delos}, and the kink in the sidechain entropy appears to be a consequence of the diminished hydrophobic compaction for large $n$ (cf. Figs. \ref{micella_ene}b and c). For the sake of simplicity we approximate the scaling of $\Delta G^{solv}$ and of $TS^{side}$ by linear functions. The general expression of these quantities, once subtracted the energies of the reference state nH, lead to
\begin{eqnarray}
E^{int}_{nM}=-315.7n+532.4, \nonumber\\
\Delta G^{solv}_{nM}=30.6-18.6n,\nonumber\\
S^{side}_{nM}=1.9n-10.9\;.
\end{eqnarray}
As the size of the micelles increases, an empty volume is created in their interior. The free energy $\Delta G^{cav}$ needed to create this cavity is, for cavities not too large with respect to the size of the solvent molecules, purely entropic and can be related to the probability  that density fluctuations produce an empty volume of the same size in bulk solvent. Computer simulations \cite{chandler} have shown that the probability distribution for such fluctuations is $\exp(-aR^2)$ with $a=58nm^{-2}$ and $R$ the radius of the cavity. Approximating the dependence of the surface of the cavity on the number of monomers as $R^2=6\cdot 10^{-2}(n-2)$ nm$^2$, one obtains
\begin{equation}
\Delta G^{cav}(n)=2.6\;T(n-2)^2
\end{equation}

The free energy of the state nM is given by the sum of all the terms discussed sofar, that is,
\begin{eqnarray}
\Delta F_{nM}&=&\Delta E_{nm}+E_{nM}^{int}+\Delta G^{solv}_{nM}-T\Delta S_{nM}^{side}-T\Delta S_{nM}^{chain}+TS^{trasl}+\Delta G^{cav}\nonumber\\
&=&49.6n-315.7n+532.4+30.6-18.6n-1.9Tn+\nonumber\\
&+&10.9T-8.8Tn+48.3T+9.6T(n-1)+2.6T(n-2)^2=\nonumber\\
&=&2.6T(n-2)^2+(-1.1T-284.7)n+563.0+49.6T.
\end{eqnarray}

Other potentially interesting states are nB, composed of non--interacting monomers in the state 1B, and nU, a disorder clamp of $n$ monomers. The free energy of the state nB is just $n$ times the free energy of the state 1B, that is
\begin{equation}
\Delta F_{nB}=-21.1n,
\end{equation}
while that of state nU can be found from the random energy model, using again Eq. (\ref{pippo}), substituting $N_c$ by the quantity $nN_c$. To be noted that, according to the present calculations, the state analyzed in ref. \cite{nussinov} gives rise to an oligomer whose free energy is $-14.5$, much higher than the free energies associated to the states displayed in Fig. \ref{fig_oligomers}, and consequently will be neglected in the following.

The summary of the free energies associated with the different states and the different sizes is displayed in Fig. \ref{f_summa} for the case of $T=295$K. The state nP$_4$ with large $n$ displays the lowest free energy minimum, while the state nM appears as a metastable state for small $n$. 

In a solution of a number $N$ of monomers, the state of each monomer can be described as a point in a two--dimensional system. The first coordinate of a point, $k$, indicates the kind of structure in which the monomer is embedded (i.e., micelle (M), disordered clump (U), etc.), while the second coordinate $l_k$ indicates the size of that structure (i.e., a micelle composed of 2, 3, ... monomers, etc.). The state of the whole system is fully determined by the coordinates $\{k,l_k\}$ of each monomer.

All the calculations performed in the present Section can be summarized in the free energy density $\zeta(k,l_k)$, which is the free energy associated with a single monomer embedded in a state of kind $k$ composed, overall, of $l_k$ monomers. This quantity reads
\begin{equation}
\label{dfree}
\zeta(k,l_k)=\left\{  
\begin{array}{ll}
f_k+\frac{s_k}{l_k} & \mbox{if $k\neq M$}\\
\frac{h_M}{l_M}(l_M-2)^2+f_M+\frac{s_M}{l_M} & \mbox{if $k=M$},
\end{array}
\right.
\end{equation}
where $f_k=\epsilon^{mon}(k)+\epsilon^{int}(k)-T\sigma^{side}(k)-T\sigma^{chain}(k)+\rho^{solv}(k)+\sigma^{trasl}$ contains the contribution proportional to the size of the aggregate, $s_k=\epsilon^{int}_0-T\sigma^{trasl}_0$ contains the contribution proportional to the number of different aggregates and $h_M=2.6T$ is the factor associated to the quadratic contribution of the micelle (state "M").

\subsection{Population of aggregates}

From the knowledge of the free energies associated with the different states of aggregation, it is possible to study how a number $N$ of monomers occupy the different states. In doing this, we will neglect the interaction between different aggregates. The study of this kind of interaction is certainly an interesting problem, but, involving the diffusion of large objects, takes place on a time scale which is longer than that of monomers and oligomers.

In order to use the expression we found for the free energies to obtain information about the thermodynamics of a solution of peptides monomers, we employ the occupation--number formalism which is often used to describe quantum many--body systems. In other words, the state of the whole system is characterized by the set of numbers $\{n(k,l_k)\}$ which indicate how many of the $N$ monomers are in the state $k$ of size $l_k$. The total free energy of the system is then
\begin{equation}
F(\{n(k,l_k)\})=\sum_{k}\sum_{l_k}^{+\infty}n(k,l_k)\zeta(k,l_k),
\end{equation}
where the sum is performed over all possible states of aggregation and all possible sizes of the aggregates, and $\zeta(k,l_k)$ is the associate free energy density, as found in Eq. \ref{dfree}.

It is convenient to calculate the partition fuction in the grand canonical ensemble, since a constrain in the number of monomers is difficult to handle analitically, which gives
\begin{equation}
{\cal Z}=\sum_{\{n(k,l_k)\}}\frac{1}{n(1,1)!\,n(1,2)!...n(k,l_k)!}\exp\left[-\beta\sum_{k}\sum_{l_k}^{+\infty}n(k,l_k)\left(\zeta(k,l_k)-\mu\right)\right],
\end{equation}
where $\beta$ is the inverse temperature and $\mu$ is the chemical potential, responsible for setting the average number of monomers in the system. The factor before the exponential takes into account the fact that monomers are indistinguishable, and consequently the swap of two of them does not produce a new state of the system. The partition function can be seen as a geometric series, which can be summed up giving
\begin{equation}
{\cal Z}=\prod_{k,l_k}   \exp\left[ \exp\left( -\beta(\zeta(k,l_k)-\mu) \right) \right].
\end{equation}
The grancanonical potential $\phi$ can be found as $-\beta^{-1}\log{\cal Z}$ and, substituting the expression of $\zeta(k,l_k)$ given in Eq. \ref{dfree}, gives
\begin{eqnarray}
\phi(\beta,\mu)&=&-\beta^{-1}\sum_{k\neq M,l_k}\exp\left[-\beta\left(f_k+\frac{s_k}{l_k}-\mu\right)\right]-\nonumber\\
&-&\beta^{-1}\sum_{l_M}\exp\left[-\beta\left(\frac{h_M}{l_M}(l_M-2)^2+f_M+\frac{s_M}{l_M}-\mu\right)\right].
\end{eqnarray}
One can then obtain the average number of monomers occupying a given state $k$ from
\begin{equation}
\label{nk}
<n_k>=\frac{\partial\phi}{\partial f_k}=\left\{         
\begin{array}{ll}
e^{-\beta(f_k-\mu)}\left[Ne^{-\beta s_k/N}+\beta s_k Ei\left(-\frac{\beta s_k}{N}\right)\right] & \mbox{if $k\neq M$}\\
\exp\left[-\beta\left(f_M-4h_M-\mu+2h_M\left(4+\frac{s_M}{h_M}\right)^{1/2} \right)\right]  & \mbox{if $k=M$,}
\end{array}
\right.
\end{equation}
where one has summed the second expression by saddle point technique. Performing the higher derivatives, one could obtain all the moments of the distribution. In order to obtain the value of the chemical potential $\mu$ for a given total number of monomers $N$ (which one usually knows, having prepared the solution at a given concentration), we set $\sum_k <n_k>=N$, obtaining
\begin{equation}
\label{mu}
\mu=\beta^{-1}\log\frac{N}{e^{-\beta(f_M-4h_M+2h_M(4+s_M/h_M)^{1/2}}+\sum_k (Ne^{-\beta s_k/N}+\beta s_k Ei(-\beta s_k/N))},
\end{equation}
being $Ei$ the exponential integral $Ei(x)\equiv -\int^{\infty}_{-x}dt\;e^{-t}/t$.

It is now possible to describe the thermodynamics of a solution of $N$ interacting monomers. Using Eq. (\ref{nk}) together with the chemical potential found in Eq. (\ref{mu}), it is possible to define a probability for a monomer to be in state $k$ as the average number of monomers in the state $k$ over the total number of monomers, namely $p(k)=<n_k>/N$.

The chemical potential $\mu$ as a function of the total number of particles of the solution is shown in Fig. \ref{fig_aggreg}(a), for temperature $T=2.5$ (equal to $295$K) and $T=3.5$ (equal to $420$K). The condition of local stability requires that $\partial\mu/\partial N\geq 0$, condition which is satisfied, for both temperatures, for $N\lesssim 100$ or for $N\rightarrow\infty$. Under the condition $N\lesssim 100$ the equilibrium state is the micelle (see Fig. \ref{fig_aggreg}(b) ), while above that number of monomers the state nP$_4$ becomes overwhelming, the other states palying essentially no role. As it is clear from Figs. \ref{fig_aggreg}(c) and (d), the population of the states nM and $nP_4$ depends weakly on the temperature.

It is worth to highlight the fact that, although the free energy of micelles is approximately linear with their length, it displays a nontrivial behaviour like the crossover shown in Fig. 7(b) and (c). The reason for this behaviour is associated with the fact that the probability of the state nP$_4$ depends not only on the associated free energy $\Delta F_{nP4}$, but also on the partition function, and consequently on the free energy of all other states, including that of micelles, which is nonlinear in the number of monomers which builds out the structure. This fact emphasises the important role  micelles have on the behaviour of the system.

\section{Discussion}

The unusual features of the peptide A$\beta$--amyloid 12--28, which encodes for an alpha--helical structure when embedded in the whole protein, while it folds to a well-defined beta--hairpin when cleaved, are related to the specific chemistry of the peptide and are discussed in detail elsewhere \cite{fabio}. 
The thermodynamics of the isolated peptide is non trivial, in that, in the range of biological temperatures, it gets more structured, the higher is the temperature. This is clear from Fig. \ref{mathe}, which displays the fractional population of the $\beta$--hairpin state 1B growing as function of the temperature up to $\approx 315$K. This behaviour is due to the fact that the stability of the state 1B, unlike the stability of its competitors (1H and 1L), relies heavily on its chain entropy. The higher the temperature is, the more important the entropic term becomes. This behaviour has been measured through circular dichroism  \cite{serpell}. In this reference the authors report percentages of $\beta$--structures of 30\% at $278$K, of 50\% at $308$K and 70\% at $328$K.

At temperatures higher than $315$K the unfolded state 1U, which is also entropically stabilized, gets importance, and consequently the population of the state 1B starts decreasing. At high temperatures the state 1U becomes dominant and the peptides gets completely unstructured. Note that, anyway, the predictions of the present model cannot go further than the range of biological temperatures, because the solvent phase transitions which take place at higher temperatures make any extrapolation of the free energy meaningless.

The present model also indicates that a single infinite--long fibril of kind nP$_4$ is the true ground state of the system. This is not consistent with atomic force microscopy experiments, which describe a population of fibrils of heterogeneous length \cite{blackley}. The reason for this result is quite simple. The stabilization energy of small oligomers is of the order of $T$, while that of fibrils is of the order of $10^2-10^3T$, and consequently, once they are formed, they are not able to separate. Moreover, it is reasonable to assume that, as fibrils grow, their diffusion coefficient decreases. As a consequence, if several incipient fibrils are present at a given time, they cannot but grow, competing for the available monomers in solution, until they have completely depleted the solution. The result of such a process is a number of fibrils of heterogeneous length, which are "frozen" in the sense that they represent an out-of-equilibrium state strongly dependent on the initial condition (it is a case of "kinetic partitioning"). An important consequence of this fact is that most likely the system will never reach its thermodynamical equilibrium, while the experimentally observed states will be metastable states, determined by the initial events of the oligomerization process.

Generalizing this idea, one could speculate that also the inner structure of fibrils is dependent on the initial stages of the dynamics. In fact, in the literature are described experiments where $\beta$--amyloid peptides of different length, but all including the central hydrophobic region, aggregate in fibrils \cite{foot6} of different diameters and different section shape. In ref. \cite{benzinger} peptide $\beta$A10--35 is prepared at 273K, concentration 0.2 mM and pH 5.6 or 7.4, and fibrils of diameter $80--90\AA$ are observed. In ref. \cite{serpell} the peptide $\beta$A11--25 is prepared at 6 mM concentration, while temperature and pH are not specified. They observe fibrils displaying an elongated section whose longer axis is $30\AA$. The peptide $\beta$A16--22, prepared at 299K, 1mM concentration and pH 7, builds fibrils characterized by antiparallel $\beta$--strands \cite{balbach}. 

These three kinds of fibrils observed experimentally are structurally similar to the states nP$_4$, nP$_2$ and nC, respectively, states which display a similar stabilization free energy (cf. Fig. \ref{f_summa}). Consequently, if at the initial stages an incipient fibril of a given type (e.g., nP$_4$, nP$_2$ or nC) is favoured, that kind will become dominant even if it does not correspond to the global free energy minimum. Within this context, the abundance of states 1B among monomeric states at high temperature could explain the uprise of fibrils of kind nC described in ref. \cite{balbach}, in that these monomers can assemble into nC fibrils without encountering any major free energy barrier. The process is then driven by diffusion, and the time needed to start fibrils is proportional to the square concentration of free monomers. At lower temperatures, as in the case of the experiment described in ref. \cite{benzinger}, the monomeric peptides that could be in solution are not in the 1B state, and consequently the fibrillation mechanism could follow another pathway, involving more complex elementary units, namely micelles, and leading to fibrils with a larger section.

If fibrillogenesis is strongly dependent on the initial conditions, then micelles will play an important role. In fact, micelles are local equilibrium states which are stable when the number of monomers in the system is low. At the initial stages of the dynamics (after $t$ seconds), the effective volume that each monomer can span is $V_{eff}=(Dt)^{3/2}$, where $D$ is the diffusion coefficient, of the order of $10^{-10}\;m^2/s$ \cite{jarvet}. Consequently, the effective number of monomers which is relevant for the out--of--equilibrium thermodynamics is $N_{eff}=c(Dt)^{3/2}$, where $c$ is the concentration. For a concentration $c=100\;\mu$M one obtains $N_{eff}=100$, compatible with the stability of micelles (cf. Fig. \ref{fig_aggreg}), one obtains $t=10^{-3}$s. This means that if incipient fibrils are not created in the first millisecond from the direct assembly of monomers, the free monomers and oligomers will aggregate into micelles, whose size is of the order of $l_M=(4+s_M/h_M)^{1/2}\approx 15$ (cf. Eq. (\ref{nk})). This is consistent with the results obtained by Benedek and coworkers by small angle neutron scattering experiments \cite{benedek1,benedek2} and with the model they propose, although for the whole peptide A$\beta$1--40. Once micelles are formed, they are remarkably stable (see Fig. \ref{f_summa}).

The important role of micelles is to deplete free monomers and small oligomers from the solution, which otherwise could build thin fibrils (of kind nC and nP$_1$). Fibrils are then cretated not directly from monomers, but as transformation of micelles, as suggested in ref. \cite{benedek1,benedek2}. The free energy barrier to go from a micelle to a fibril (e.g., of kind nP$_4$) describes the internal rearrangement of monomers and cannot be derived by the present model. Anyway, it suggests that the number of monomers in fibril seeds, that is the shortest stable fibril, is of the order of $l_M\approx 15$.

Micelles are then a finite--size effect which plays a fundamental role in the fibrillogenesis mechanism. This turns out not to be a real nucleation process. In fact, we suggest that fibrils arise either by spontaneous assembly of monomers without any major free energy barrier, or by transformation of a micelle--like metastable state into a fibril of equal length, depending on the conditions of the solution.

\section{Conclusions}

We make use of MD simulations to analyze the mechanism which makes A$\beta$--fragments aggregate into fibrils. Since a brute--force MD approach is neither computationally feasible, nor leads to a real insight into the aggregation thermodynamics, we use MD trajectories as a starting point in order to detect those states which are thermodynamically important and to characterize their free energy. Making use of these informations, we build a simple physical model which teaches us some important facts about the A$\beta$--fragments: 1) in the range of biological temperatures, the peptides is more stabilized into a $\beta$ structure, the higher is the temperature; 2) micelles are an important metastable state which is populated at the beginning of the dynamical process and subtract monomers and small oligomers from the solution; 3) fibrils arise either by direct assembly of the few free monomers (at relatively high temperature) or by transformation of a micelle into a fibril of the same size.



\appendix
\section{Estimate for the rotational entropy}

If one of the monomers is selected to set the reference axes, each of the $N_m-1$ other monomers, when not bound, can rotate freely in space, the product the amplitudes of its three Euler angles being $4\pi^3$. When bound to another monomer, the three angles are constrained. To evaluate to which extent they are constrained, we calculate the average vibration length of the bond distances between two monomers and obtain the angular amplitude by trigonometric transformations. For example, the length of the longer axis of the 1B conformation is $\approx 20\AA$ and the average vibration length of the intra--monomer bonds is $2\AA$. Consequently, the angular amplitude associated to the vibration of the two bound monomers along their longer axis is $\tan^{-1}(2/20)=0.1$. Similarly, one finds for the other two angular amplitudes the values $0.4$ and $0.2$ (assuming that the width of the 1B conformation is $\approx 5\AA$). The product of the three angular amplitudes give $\Delta\omega=8\cdot 10^{-3}$.



\newpage


\begin{figure}[h!]
\caption{The conformation of some relevant monomeric states.}
\label{fig_dimers}
\end{figure}

\begin{figure}[h!]
\caption{The conformation of some relevant oligomeric states. Each state is built out of a number $n$ of monomers, each of them being in the state 1B. States nP$_1$, nP$_2$, nP$_4$ and nC display translational symmetry in the direction perpendicular to the plane defined by the hairpin, while state nM displays discrete rotational symmetry.}
\label{fig_oligomers}
\end{figure}

\begin{figure}[h!]
\caption{A sketch of the backbone degrees of freedom associated with the state 1B.}
\label{fig3new2}
\end{figure}

\begin{figure}[h!]
\caption{The relative population of state 1B as a function of temperature. In the inset, the dependence of the free energy of the state 1U as a function of T.}
\label{mathe}
\end{figure}

\begin{figure}[h!]
\caption{The internal energy (a), the total solvatation free energy (b) and the total energy associated with the sidechain entropy (c) of oligomeric micelles, as a function of the number of monomers. These quantities are found from MD simulations at 295 K.}
\label{micella_ene}
\end{figure}

\begin{figure}[h!]
\caption{The free energies associated with the different aggregation states at $T=295$K, displayed as a function of the size of the aggregate.}
\label{f_summa}
\end{figure}

\begin{figure}[h!]
\caption{(a) The chemical potential as function of the total number $N$ of particles at temperatures T=2.5 kJ/mol ($=295$K) and T=3.5 kJ/mol ($=420$K). (b) The fractional population of states nP$_4$ and nM at T=2.5 kJ/mol as a function of $N$. (c) and (d) The fractional population of states nP$_4$ and nM as a function of temperature, for $N=100$ and $N=10000$, respectively.}
\label{fig_aggreg}
\end{figure}

\clearpage
\newpage

\begin{table}
\begin{tabular}{|c|c|c|c|c|c|}\hline
State & $\Delta E$ & $T\Delta S^{side}$ & $\Delta G^{solv}$ & $T\Delta S^{chain}$ & $\Delta F$\\\hline
1H & 0 & 0 & 0 & 0 & 0 \\
1B & 5.9 & -11.4 & -16.4 & 21.6 & -20.8\\
1L & -168.7 & -148.8 & 3.9 & 0 & -16.0\\
1U & 48.4 & 2.0 & -19.4 & 39.3& -12.3 \\\hline
\end{tabular}
\caption{The energy terms associated with monomeric states.}
\label{newtab1}
\end{table}

\begin{table}
\begin{tabular}{|c|c|c|c|c|c|c|}\hline
State & $\Delta E$ & $E^{int}$ & $TS^{side}$ & $TS^{chain}$ &$\Delta G^{sol}$ & $\Delta F'$ \\\hline
2P$_1$ & -62.0 & -135.9 & -8.8 & 0 & -9.0 & -198.1 \\
2P$_2$ & -37.6 & -62.7 & -0.8 & 44.2 & -24.8 & -168.5 \\
2C & -23.6 & -152.2 & +9.4 & 0 & -22.6 & -207.8 \\
2K & +58 & -124.1 & -32.8 & 0 & -15.6 & -49.2 \\
2M & +14 & -140.0 & -16.1 & 44.2 & -30.5 & -184.9  \\\hline
4P$_4$ & +7.2 & -153.8 &  +12.8 & 0 & -70.2 & -229.6 \\
4P$_1$ & -128.0 & -293.6 &  +4.8 & 0 & -20.7 & -446.5 \\
4M & +96.0 & -785.2 & -38.8 & 88.4 & -15.9 & -754.7 \\\hline
8M & +397.6 & -1975.2 & 102.4 & 176.8 & -127.0 & -1983.8\\
8P$_4$ & -150.4 & -801.6 & -18.4 & 0 & -135.4 & -567.4 \\\hline
\end{tabular}
\caption{The free energies of relevant aggregated states with reference to that state nH. The results were obtained from MD simulations at 295 K. The quantity $\Delta F'$ is the free energy of the associated state, except for the contribution of the translational entropy and of the entropy necessary to create a cavity within the micelles.}
\label{table_2f}
\end{table}

\begin{table}
\begin{tabular}{|c|c|c|c|c|c|c|}\hline
State & $\epsilon^{mon}$ & $\epsilon^{int}$ & $\epsilon^{int}_0$ & $\sigma^{side}$ & $\sigma^{chain}$ & $\rho^{solv}$\\\hline
nP$_1$ & -31.0 & -135.9 & +135.9 & -1.8 & 0 & -4.5 \\
nP$_2$ & -18.8 & -128.9$^*$ & +195.2$^*$ & -0.16 & 0 & -12.4 \\
nP$_4$ & -18.8 & -161.9 & +481.0 & -0.16 & 0 & -16.9 \\
nC & -11.8 & -152.2 & +152.2 & 1.9 & 0 & -11.3 \\\hline
\end{tabular}
\caption{The energy parameters which generalize the results listed in Table 2 (cf. Eq. (\protect\ref{dfree})). ($^*$) is found from the combination of states 2P$_2$ and 4P$_1$.}
\label{free_dens}
\end{table}


\begin{thebibliography}{99}
\bibitem{foot1} In 1853, the German pathologist Rudolf Virckow described eosinophilic waxy tissue deposits which he called amyloid (starch like). The misnomer has proven resilient, although all known amyloids are composed mainly of proteins.
\bibitem{landsbury} P. T. Landsbury, Proc. Natl. Acad. Sci. USA {\bf 96}, 3342 (1999)
\bibitem{goldberg} M. S. Goldberg and P. T. Landsbury, Nature Cell. Biol. {\bf 2}, E115 (2000
\bibitem{lashuel} H. A. Lashuel, D. Hartley, T. Waltz and P. T. Landsbury, Nature {\bf 418}, 291 (2002)
\bibitem{hartley} D. M. Hartley {\it et al.}, J. Neurosci. {\bf 19} 8876 (1999)
\bibitem{lambert} M. P. Lambert {\it et al.} Proc. Natl. Acad. Sci. USA {\bf 95}, 6448 (1998)
\bibitem{walsh} D. M. Walsh {\it et al.}, Nature {\bf 416}, 535 (2002)
\bibitem{kirschner} D. A. Kirschner, C. Abraham and D. J. Selkoe, Proc. Natl. Acad. Sci. USA {\bf 83}, 503
\bibitem{talafous} J. Talafous, K. J. Marcinowski, G. Klopman and M. G. Zagorski, Biochemistry {\bf 33}, 7788 (1994)
\bibitem{serpell} L. C. Serpell, C. C. F. Blake and P. E. Fraser, Biochemistry {\bf 39}, 13269 (2000)
\bibitem{harper} J. D. Harper, S. S. Wong, C. M. Lieber and P. T. Lansbury, Biochemistry {\bf 38}, 8972 (1999)
\bibitem{gromacs} H, J. C. Berendsen, D. van der Spoel and R. van Drunen, Comp. Phys. Comm. {\bf 91} 43 (1995)
\bibitem{jchemphys2} R. A. Broglia and G. Tiana, J. Chem. Phys. {\bf 114}, 7267 (2001)
\bibitem{jchemphys1} G. Tiana, R. A. Broglia, H. E. Roman, E. Vigezzi and E. I. Shakhnovich, J. Chem. Phys. {\bf 108}, 757 (1998)
\bibitem{jchemphys3} G. Tiana and R. A. Broglia, J. Chem. Phys. {\bf 114} 2503 (2001) 
\bibitem{aggreg} R.A. Broglia, G. Tiana, S. Pasquali, H. E. Roman, E. Vigezzi, Folding and Aggregation of Designed Protein Chains, Proc. Natl. Acad. Sci. USA, {\bf 95} 12930 (1998)
\bibitem{foot2} In what follows, all energies are expressed in kJ/mol.
\bibitem{fabio} F. Simona, G. Tiana, R. A. Broglia and G. Colombo (to be published)
\bibitem{foot3} To be noted that because $pV$ is essentially constant in our calculations, one can use both Helmoltz ($F=E-TS$) as well as Gibbs ($G=E+pV-TS$) free energy.
\bibitem{foldx} R. Guerois, J. E. Nielsen and L. Serrano, J. Mol. Biol. {\bf 320}, 369 (2002)
\bibitem{derrida} B. Derrida, Phys. Rev. B, {\bf 24}, 2613 (1981)
\bibitem{foot4} I.e., the distribution of RMSD to the state 1B shows a bimodal shape: we define as structurally different those conformations belonging to the peak centered about the larger value of RMSD.
\bibitem{foot5} That is $\epsilon_0=-31.2$ kJ/mol. This quantity, which corresponds to an energy of -7.5 kcal/mol, equivalent to $\approx 3.8\cdot 10^3$K is very large on the basis of the marginal stability displayed by proteins under biological conditions. To be noted that $\epsilon_0$ contains not only the interaction between amino acids, but also the internal energy arising from the interaction between all the atoms which constitute each amino acid.
\bibitem{creighton} T. E. Creighton, {\it Proteins}, W. H. Freeman and Co., New York (1993)
\bibitem{delos} G. Caldarelli and P. De Los Rios, J. Biol. Phys. {\bf 27}, 229 (2001)
\bibitem{chandler} D. M. Huang and D. Chandler, Phys. Rev. E. {\bf 61}, 1501 (2000)
\bibitem{nussinov} B. Ma and R. Nussinov, Proc. Natl. Acad. Sci. USA, {\bf 99}, 14126 (2002)
\bibitem{blackley} H. K. L. Blackley {\it et al.}, J. Mol. Biol. {\bf 298}, 833 (2000)
\bibitem{foot6} To be precise, "protofibrils", since usually the term "fibril" is used for thicker aggregates which are though to be produce by the pairing of more protofibrils.
\bibitem{benzinger} T. L. S. Benzinger {\it et al.}, Biochemistry {\bf 39}, 3491 (2000)
\bibitem{balbach} J. J. Balbach {\it et al.}, Biochemistry {\bf 39}, 13748 (2000)
\bibitem{jarvet} J. Jarvet, P. Damberg, K. Bodell, L. E. G. Eriksson and A. Gr\"aslund, J. Am. Chem. Soc. {\bf 122}, 4261 (2000)
\bibitem{benedek1} A. Lomakin, D. S. Chung, G. B. Benedek, D. A. Kirschner and D. B. Teplow, Proc. Natl. Acad. Sci. USA {\bf 93}, 1125 (1996)
\bibitem{benedek2} W. Yong, A. Lomakin, M. D. Kirkitadze, D. B. Teplow, S. H. Chen and G. B. Benedek, Proc. Natl. Acad. Sci. USA {\bf 99}, 150 (2002)
\end{thebibliography}
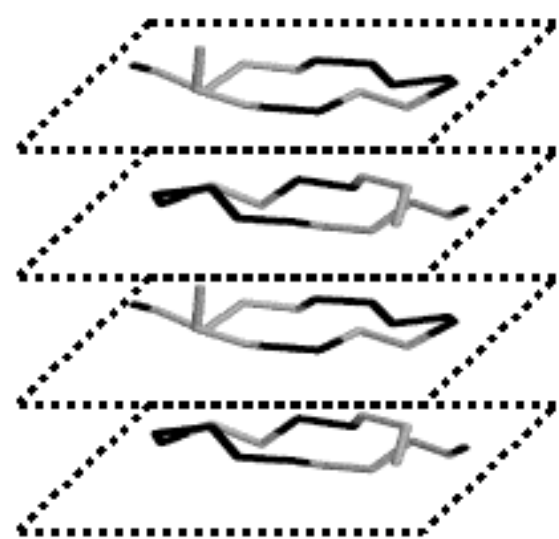
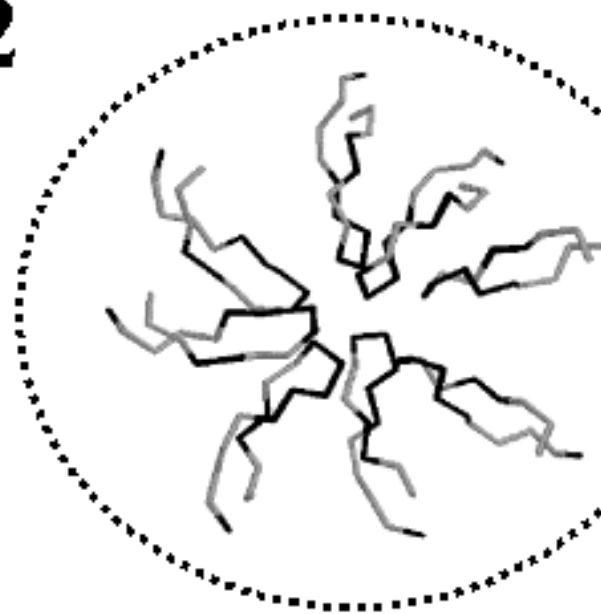
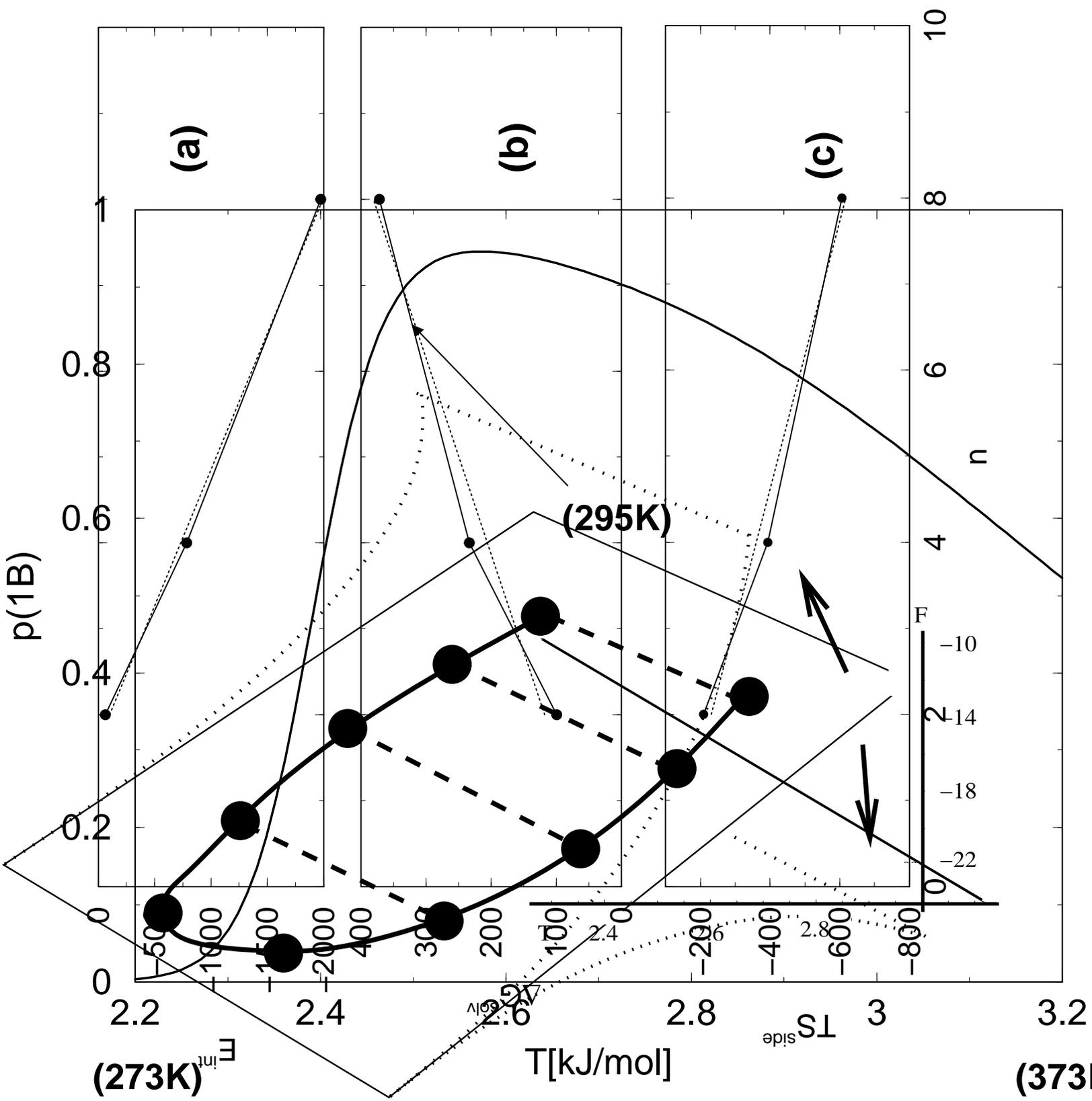
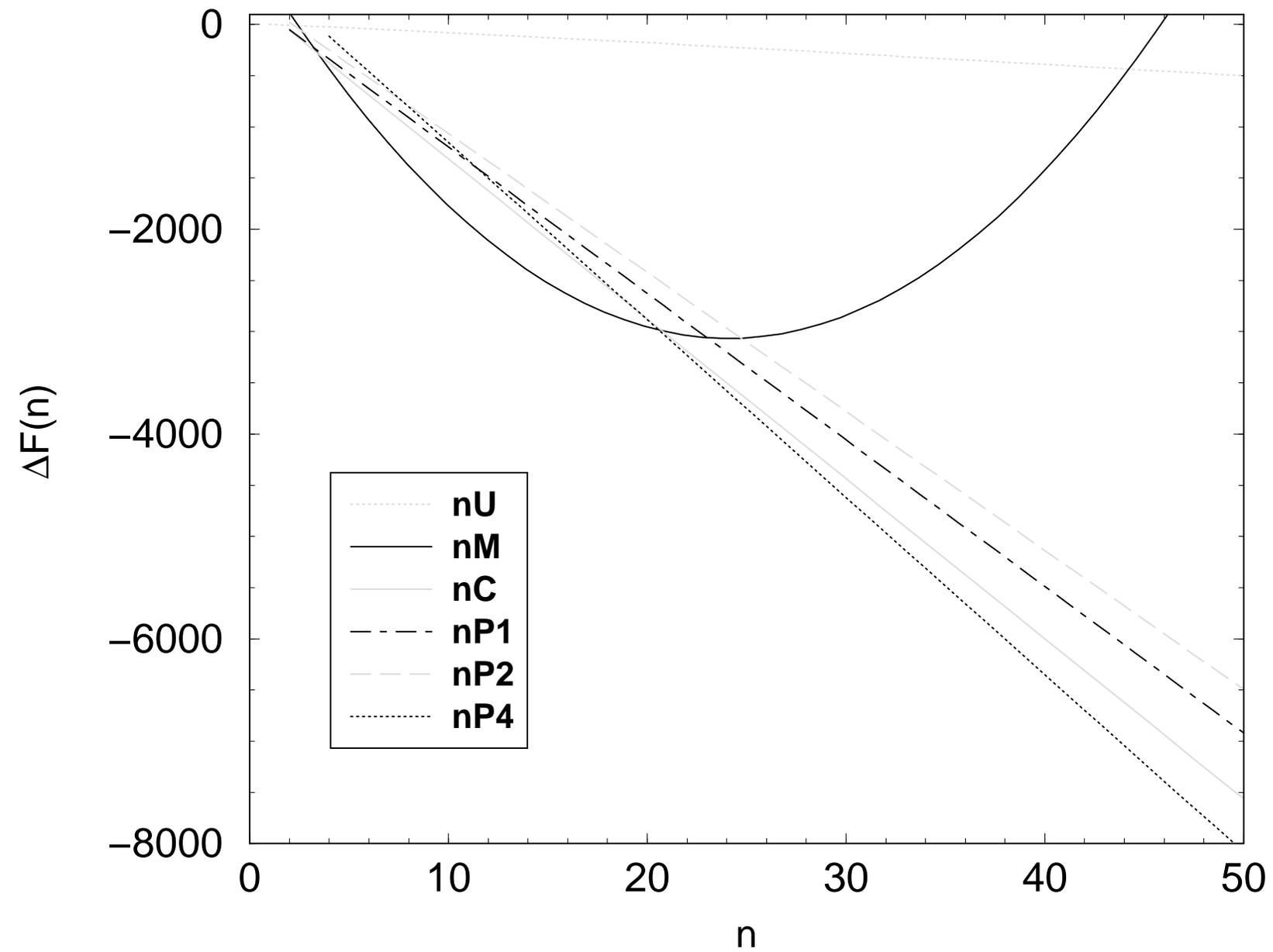
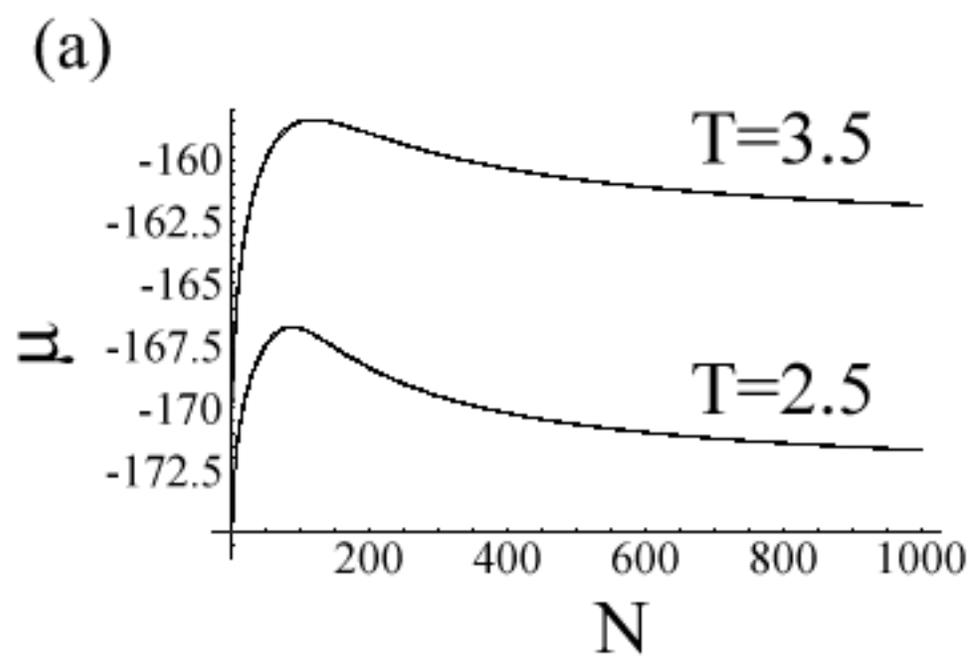
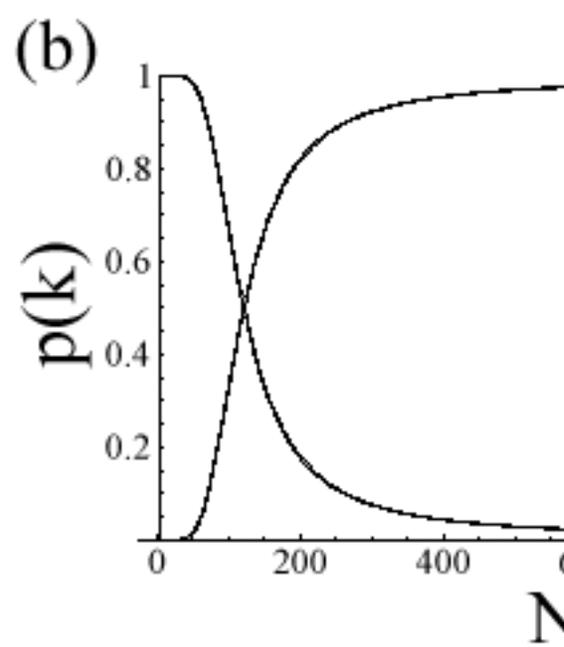
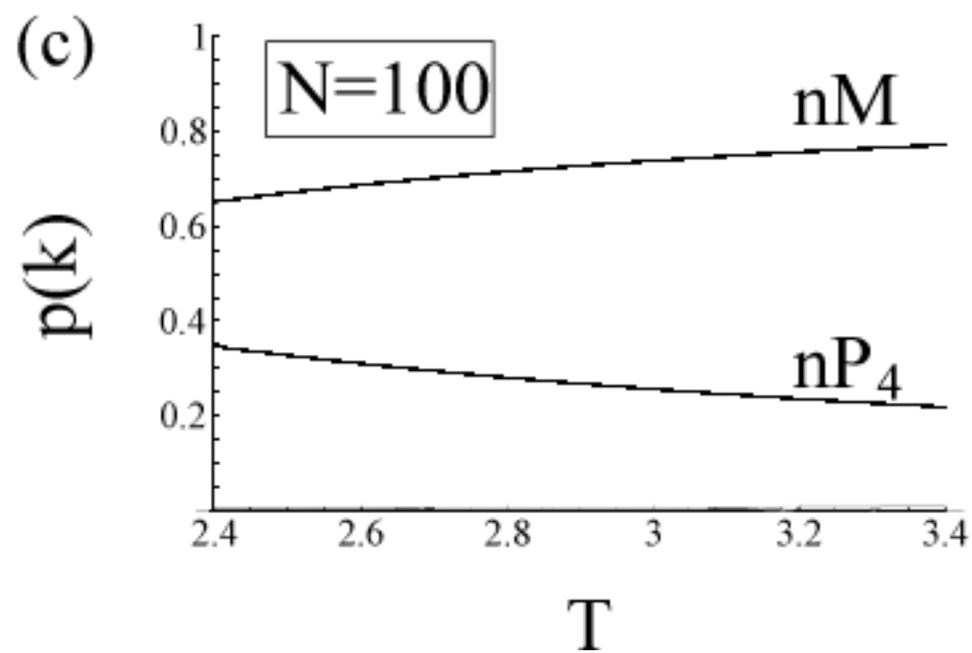
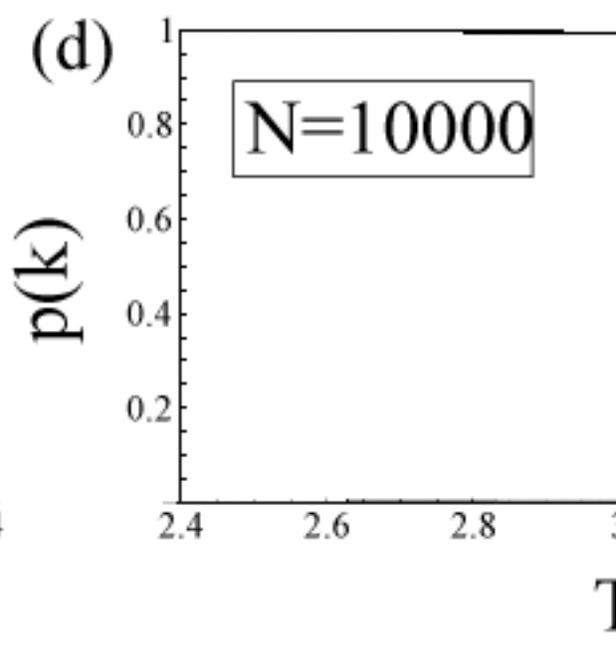
\end{document}